# Continuous Three-level Quantum Heat Engine with High Performance Under Medium Temperature Difference

Gao-xiang Deng[a], Wei Shao[a,b], Yu Liu[a], Zheng Cui[b,*]

a Institute of Thermal Science and Technology, Shandong University, Jinan, 250061, P.R. China

b Shandong Institute of Advanced Technology, Jinan, 250100, P.R. China

*Correspondence: Tel. & Fax: 86-88399000-2511; Email: zhengc@sdu.edu.cn;

## Abstract

The possibility of utilizing quantum effects to enhance the performance of quantum heat engines has been an active topic of research, but how to enhance the performance by optimizing the engine parameters needs to be further studied. In this study, the temperature difference and dissipation modes affecting the performance of a three-level quantum heat engine were analyzed using an orthogonal test. The results indicated that the dissipation mode dominated the performance of the quantum heat engine. The quantum heat engine performs best when there is only resonance and no detuning; however, when detuning exists, a lower resonance can improve the efficiency by reducing energy losses. Regarding the temperature difference, the best performance was achieved at medium temperature difference owing to the decreasing heat leakage. Finally, the "quantum friction" caused by the detuning could make the maximal efficiency lower than the Carnot efficiency.

**Keywords:** Quantum heat engine; Continuous Coupling; Performance optimization; Orthogonal test; Quantum friction.

# 1 Introduction

A classical heat engine generally consists of three parts: hot reservoir, cold reservoir, and working fluid. The hot and cold reservoirs are sometimes referred to as environments without distinction. The working fluid is connected to the environment to exchange heat and output work. Scovil and Schulz-DuBois [1, 2] first introduced the concept of a quantum heat engine (QHE) decades ago. Currently, the QHE generally refers to small engines with quantum effects that cannot be described by classical thermodynamics, such as the coherent state of the working fluid or entanglement between the working fluid and the environment. Studies on the QHE can reveal the principles of microscopic thermodynamics and promote the development of nanotechnology [3].

One of the key questions in QHE research is whether quantum effects can enhance the performance. This issue has been debated in the literature for several decades. Initially, with the analysis of the quantum Carnot cycle, which is an extended version of the classical Carnot cycle to the quantum regime, it was thought that the quantum effects of the QHE cannot enhance efficiency beyond the Carnot limit [4-8]. It is generally even worse than the Carnot efficiency owing to the "quantum friction" [9-12] that consumes the energy that was supposed to be extracted as output work, resulting in a lower efficiency. To obtain analyzable solutions, most of the above analyses were based on the assumptions of an initial Gibbs state and a sufficiently thermalized process for the QHE. Many scholars believe that the above assumption leads to the neglect of quantum effects therein [13-18]. It was found that nonthermal resources, such as the initial coherent state [13-15] or squeezed thermal bath [16-18], can be used to improve the performance of the QHE. Unfortunately, to obtain these resources, energy is consumed, and as a result, there is no performance improvement [7, 19-22]. More recently, however, theoretical studies have indicated that the quantum effects (e.g., coherence) generated during operation, rather than those previously produced, can be used to improve the performance of the QHE [23-25]. For example, Uzdin et al. [23] theoretically demonstrated that the internal coherent superposition states created during the operation of a QHE affect its performance.

To answer this question, that is, whether the quantum effects within the QHE can be used to enhance its performance, many experiments on the QHE have been implemented in recent years [26-43] and have provided some positive answers [31-33, 40]. For example, Ji et al. [33] used lasers to control the nitrogen atoms and electrons of negatively charged nitrogen vacancy (NV⁻) centers in diamond as the working fluid and environment, respectively. They then built a quantum Szilard heat engine with CROT gates acting as "demon" and showed that the entanglement between nitrogen atoms and electrons improved the performance of the QHE. In addition, Klatzow et al. [31] built a QHE with NV⁻ centers in diamond and showed that when the thermal processing time was less than the decoherence time, the coherence generated during operation could be used to enhance the performance of the QHE.

However, the principle by which quantum effects lead to the improved performance of heat engines remains to be further explained. Recently, Bayona-Pena and Takahashi [44] established a steady-state model of a three-level QHE to investigate the effect of quantum coherence on QHE performance. They then explored the impact of different parameters on the steady-state performance of the three-level QHE and pointed out that a non-diagonal heat flux is a necessary condition for a QHE to improve its performance. Nevertheless, the cases studied in that work may be incomplete, and the impact of each parameter on performance may not be sufficiently clear.

Based on the above deficiencies, this study conducted an orthogonal test to analyze how different parameters affect the performance of the QHE more comprehensively. This paper is organized as follows. An introduction to the QHE and the previous research are provided in the first part. The second part describes the model and method of this study, specifically the model of the three-level QHE and the method of the orthogonal test. In the third part, the orthogonal test is carried out, and then the results and "quantum friction" are further analyzed. Finally, the conclusions of this study are summarized.

## 2 Model and Method

### 2.1 Model

A model of a three-level QHE is shown in Fig. 1. The quantum system (working fluid in the QHE) absorbs heat $Q_h$ from the hot reservoir, releases heat $Q_c$ to the cold reservoir, or outputs work $W$ to the external field $V$ when the system is coupled to the corresponding environment (reservoirs or field), as shown in Fig. 1(a). Fig. 1(b) shows the energy levels of the quantum system in the three-level QHE, $|0\rangle$, $|1\rangle$, and $|2\rangle$. The dynamics of the transitions and fallbacks between these energy levels are the fundamental thermal processes of a QHE. For example, the hot reservoir at temperature $T_h$ coupled to the energy-level gap between $|0\rangle$ and $|2\rangle$ can excite a transition from $|0\rangle$ to $|2\rangle$ and lead to the heat absorption of $Q_h$. However, the cold reservoir at temperature $T_c$ coupled with the gap between $|0\rangle$ and $|1\rangle$ can cause a fallback from $|1\rangle$ to $|0\rangle$ and the heat release of $Q_c$. The external field $V$ coupled to the gap between $|1\rangle$ and $|2\rangle$ can be used to extract work $W$.

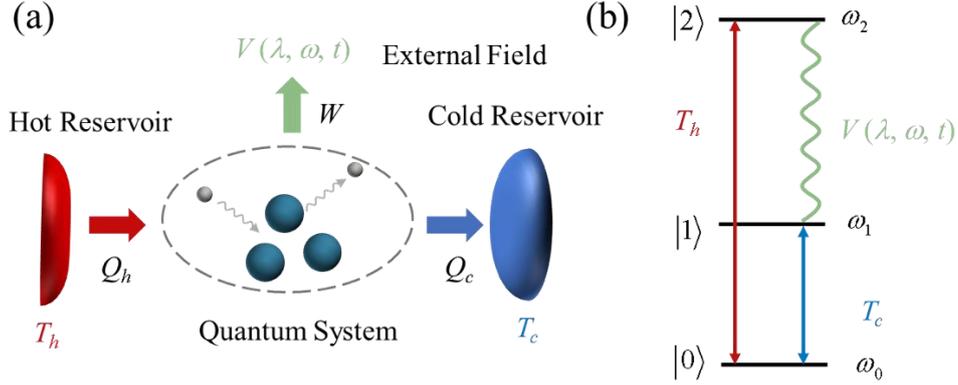

Fig. 1. Schematic of a three-level quantum heat engine (QHE). (a) The quantum system absorbs heat $Q_h$ from the hot reservoir at temperature $T_h$ or releases heat or $Q_c$ to the cold reservoir at $T_c$. And the coupling between the quantum system and the external field $V$ with intensity $\lambda$, frequency $\omega$ (which varies with time $t$) is used to extract work $W$. (b) Energy levels of the quantum system in a three-level QHE. The thermal relationships between these energy levels and heat reservoirs or external fields are described in the main text.

The thermodynamic process of a three-level QHE can be obtained by solving the Gorini-Kossakowski-Lindblad-Sudarshan (GKLS) equation [45],

$$\partial_t \hat{\rho}_S(t) = -i\left[\hat{H}_S(t), \hat{\rho}_S(t)\right] + \sum_\alpha \hat{D}_\alpha\left[\hat{\rho}_S(t)\right] \tag{1}$$

where $\hat{\rho}_S$ and $\hat{H}_S$ are the density matrix and Hamiltonian of the quantum system, respectively, and $\hat{D}_\alpha$ is the dissipation operator representing the dissipation mode. The subscript $\alpha$ is $c$ or $h$, which represents the dissipation to the hot or cold reservoir. Bayona–Pena and Takahashi solved the GKLS equation for the three-level QHE, as shown in Fig. 1, in the stationary limit and obtained expressions for the power

$$P = \varepsilon_{21} \frac{\omega^2 \sin^2\theta}{2G}\left(\frac{g_2^-}{g_2} - \frac{g_1^-}{g_1}\right)\frac{1}{Z} \tag{2}$$

and efficiency

$$\eta = \eta^{CP} \frac{\frac{1}{\eta^{CP}} P}{\frac{1}{\eta^{CP}} P + \rho_0 P_0} \quad (3)$$

In this study, $P\eta$ is the product of the two equations above, $\omega$ is the frequency of the external field, and $\theta$ is defined as

$$\tan \theta = \frac{2\lambda}{\omega_2 - \omega_1} \quad (4)$$

where $\lambda$ is the intensity of the external field. $g_1, g_2, g_1^-, g_2^-$ are defined by the resonant dissipation coefficient $\gamma_c(\varepsilon_{10})$, $\gamma_h(\varepsilon_{20})$ and detuning dissipation coefficient $\gamma_c(\varepsilon_{20})$, $\gamma_h(\varepsilon_{10})$, as follows:

$$\begin{aligned} g_1 &= \gamma_c(\varepsilon_{10}) \frac{1+\cos\theta}{2} + \gamma_h(\varepsilon_{10}) \frac{1-\cos\theta}{2} \\ g_2 &= \gamma_h(\varepsilon_{20}) \frac{1+\cos\theta}{2} + \gamma_c(\varepsilon_{20}) \frac{1-\cos\theta}{2} \\ g_1^- &= \gamma_c(-\varepsilon_{10}) \frac{1+\cos\theta}{2} + \gamma_h(-\varepsilon_{10}) \frac{1-\cos\theta}{2} \\ g_2^- &= \gamma_c(-\varepsilon_{20}) \frac{1+\cos\theta}{2} + \gamma_h(-\varepsilon_{20}) \frac{1-\cos\theta}{2} \end{aligned} \quad (5)$$

where

$$\gamma_\alpha(-\varepsilon) = e^{-\beta_\alpha \varepsilon} \gamma_\alpha(\varepsilon) \quad (6)$$

and $\beta_\alpha$ is the inversed temperature of the hot or cold reservoir

$$\beta_\alpha = \frac{1}{k_B T_\alpha} \quad (7)$$

where $k_B$ is Boltzmann's constant. Further details regarding other parameters such as $\varepsilon, G, Z, \eta^{CP}, \rho_0, P_0$ can be found in [44].

## 2.2 Method

To gain a deeper understanding of the influence of quantum effects on the performance of three-level QHEs under different environmental conditions, it is necessary to further optimize and analyze the three-level QHE. However, the performance of a three-level QHE is affected by many environmental parameters, which increase the calculation, optimization, and analysis workloads. The performance optimization of a three-level QHE can be regarded as a multi-objective optimization problem. Generally, such problems can be solved using the Lagrange method [46] or genetic algorithm [47]; however, these methods are sensitive to the initial value, and it is difficult to determine the weight of each parameter. The orthogonal test [48] allows the amount of testing to be reduced without losing representativeness and provides a ranking of the impact of each parameter on the performance. Therefore, it can be used for multifactor and multilevel analyses.

The orthogonal test consists of the following steps: (a) select the optimization objective and corresponding evaluation indexes, (b) select factors and corresponding levels, (c) design the orthogonal table and perform the test, and (d) analyze the result and choose the best combination. The performance of the QHE was selected as the optimization objective, and power $P$, efficiency $\eta$, and efficacy $P\eta$ were selected as indices to evaluate the performance. The temperature difference

$$\Delta\beta = \frac{1}{\beta_h \omega_{10}} - \frac{1}{\beta_c \omega_{10}} \tag{8}$$

represented by different combinations of cold temperature (inverse temperature) $\beta_c \omega_{10}$ and hot temperature $\beta_h \omega_{10}$ was set as factor 1. $\omega_{10}$ is the energy level difference between energy levels $|1\rangle$ and $|0\rangle$:

$$\omega_{10} = \omega_1 - \omega_0 \tag{9}$$

The resonant dissipation coefficient $D_r$ described by $\gamma_c(\varepsilon_{10})/\omega_{10}$ and $\gamma_h(\varepsilon_{20})/\omega_{10}$ was set to a factor of 2. The detuning dissipation coefficient $D_d$ is reflected by $\gamma_c(\varepsilon_{20})/\omega_{10}$ and $\gamma_h(\varepsilon_{10})/\omega_{10}$ was set to a factor of 3. These factors each had three levels, and their corresponding values are listed in Table 1. Levels 1, 2, and 3 of $\Delta\beta$ means low, medium, and high hot–cold reservoir temperature differences, respectively. Levels 1, 2, and 3 of $D_r$ or $D_d$ mean low, medium, and high resonance or detuning, respectively. For simplicity, low, medium, and high refer to levels 1, 2, and 3, respectively, in the following text; for example, low $\Delta\beta$ means $\Delta\beta$ level 1.

Table 1. Factors and levels used in the orthogonal test.

| Level | Factor | | | | | |
|---|---|---|---|---|---|---|
| | $\Delta\beta$ | | $D_r$ | | $D_d$ | |
| | $\beta_c\omega_{10}$ | $\beta_h\omega_{10}$ | $\gamma_c(\varepsilon_{10})/\omega_{10}$ | $\gamma_h(\varepsilon_{20})/\omega_{10}$ | $\gamma_c(\varepsilon_{20})/\omega_{10}$ | $\gamma_h(\varepsilon_{10})/\omega_{10}$ |
| 1 | 5 | 1 | 0.5 | 0.5 | 0 | 0 |
| 2 | 2.5 | 0.5 | 1 | 1 | 0.5 | 0.5 |
| 3 | 1 | 0.2 | 2 | 2 | 2 | 2 |

The $L_9(3^3)$ orthogonal test table was adopted because there are three factors, and each factor has three levels, as listed in Table 2. Cases 1–9 represent the order of the orthogonal test. In addition, 1, 2, and 3 in each column are Level 1, Level 2, and Level 3 of the corresponding factor, respectively. The specific values for each level are listed in Table 1. The analysis of range and variance determines the impact of each factor on the performance and the best combination of levels for the different factors. This will be discussed later in the "Results" section.

Table 2. Order of the designed orthogonal test.

| Case | $\Delta\beta$ | $D_r$ | $D_d$ |
|---|---|---|---|
| 1 | 1 | 1 | 1 |
| 2 | 1 | 2 | 3 |
| 3 | 1 | 3 | 2 |
| 4 | 2 | 1 | 3 |
| 5 | 2 | 2 | 2 |
| 6 | 2 | 3 | 1 |
| 7 | 3 | 1 | 2 |
| 8 | 3 | 2 | 1 |
| 9 | 3 | 3 | 3 |

## 3 Results

### 3.1 Validation of the calculations

To test the validity of the calculation results, the main results of Bayona-Pena and Takahashi were reproduced, and the second law of thermodynamics was verified. The reproduced results are presented in Fig. 2.

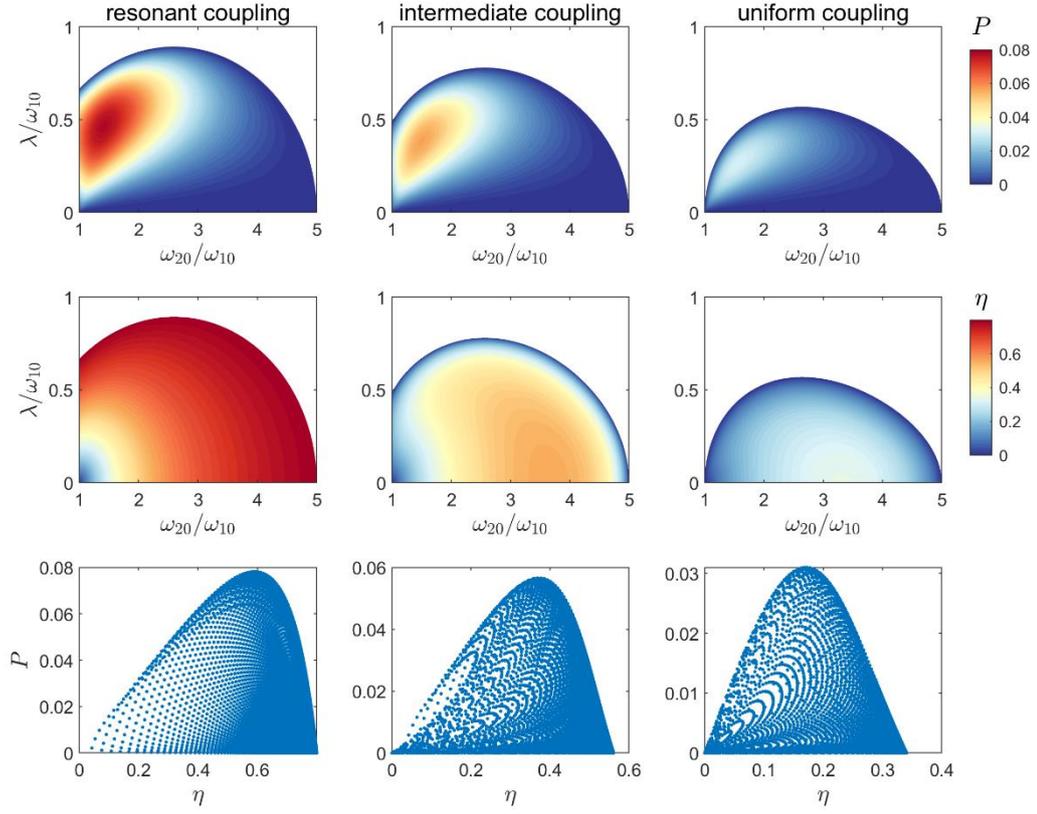

Fig. 2. Reproduced results. The plots from the left to right column correspond to different dissipation modes, i.e., resonant, intermediate, and uniform coupling. The plots from top to bottom represent the power $P$, efficiency $\eta$, and distribution $(\eta, P)$ for the different dissipation modes.

In recent years, there have been many controversies regarding the GKLS equation [49-52], particularly when solving a specific system that may violate the second law of thermodynamics [49, 51, 53-55]. Therefore, we first calculated the average entropy production of all nine cases in the orthogonal test to verify the applicability of the GKLS equation. The entropy production rate can be expressed as follows:

$$\dot{\sigma} = -\mathrm{tr}\left[\frac{\partial \hat{\rho}_S(t)}{\partial t} \ln \hat{\rho}_S(t)\right] - \sum_{\alpha=c,h} \beta_\alpha \dot{Q}_\alpha(t) \tag{10}$$

where the first term on the right-hand side is the von Neumann entropy. The average entropy production is

$$\langle \dot{\sigma} \rangle = \int \dot{\sigma}(t) dt = (\beta_c - \beta_h)\left[\left(\frac{1}{\eta^{\mathrm{CP}}} - \frac{1}{\eta^{\mathrm{C}}}\right)P + \rho_0 P_0\right] \tag{11}$$

The average entropy production of all the nine orthogonal tests was calculated using

Eq. (11) to verify the applicability of the GKLS equation: The results are presented in Fig. 3 and Table 3. The minimum average entropy production of all cases was greater than or equal to 0, which means that there was no violation of the second law of thermodynamics in the orthogonal test.

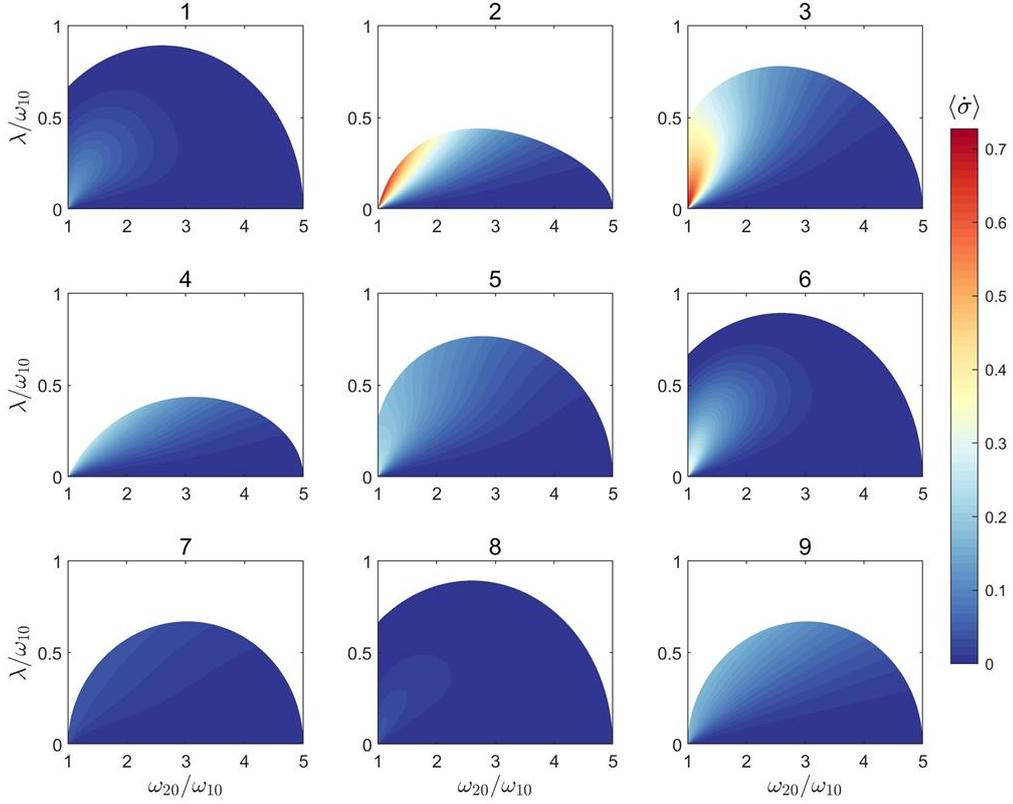

Fig. 3. Averaged entropy production $\langle\dot{\sigma}\rangle$ of each test. Each plot has the case number of the orthogonal test. The intensity of the external field $\lambda/\omega_{10}$ takes values from 0 to 1, and the energy level ratio $\omega_{20}/\omega_{10}$ ($\omega_{20} = \omega_2 - \omega_0$) takes values from 1 to 5 in each case.

Table 3 Minimum averaged entropy production of each case of the orthogonal test.

| **Case** | 1 | 2 | 3 | 4 | 5 | 6 | 7 | 8 | 9 |
| --- | --- | --- | --- | --- | --- | --- | --- | --- | --- |
| $\langle\dot{\sigma}\rangle_{min}$ | 0 | 0 | 0 | 0 | 0 | 0 | 0 | 0 | 0 |

## 3.2 Orthogonal test

The results of the orthogonal test are presented in Table 4. The intensity of the external field $\lambda/\omega_{10}$ takes values from 0 to 1, and the energy level ratio $\omega_{20}/\omega_{10}$ ($\omega_{20} = \omega_2 - \omega_0$) takes values from 1 to 5 in each case to obtain three sets of values for $P$, $\eta$, and $P\eta$, respectively. Then, the maximum values of $P$, $\eta$, and $P\eta$ are selected in the three sets as the results for each case of the test.

Table 4. Results of the orthogonal test.

| Case | $\Delta\beta$ | $D_r$ | $D_d$ | $P$ | $\eta$ | $P\eta$ |
|---|---|---|---|---|---|---|
| 1 | low | low | low | 0.023 | 0.800 | 0.014 |
| 2 | low | medium | high | 0.012 | 0.401 | 0.002 |
| 3 | low | high | medium | 0.057 | 0.561 | 0.022 |
| 4 | medium | low | high | 0.019 | 0.591 | 0.007 |
| 5 | medium | medium | medium | 0.048 | 0.671 | 0.023 |
| 6 | medium | high | low | 0.105 | 0.800 | 0.067 |
| 7 | high | low | medium | 0.024 | 0.695 | 0.013 |
| 8 | high | medium | low | 0.045 | 0.800 | 0.031 |
| 9 | high | high | high | 0.038 | 0.378 | 0.009 |

### 3.2.1 Range analysis

A range analysis determines the impact of each factor on the performance and provides the optimal level for each factor. The results of the range analysis based on Table 4 are presented in Table 5. In the first column, $K_1$, $K_2$, and $K_3$ are the sum of the results for different performances ($P$, $\eta$, or $P\eta$) of a specific factor that is taken as low, medium, and high, respectively. $\bar{K}_1$ ($\bar{K}_2$ or $\bar{K}_3$) is the result of dividing $K_1$ ($K_2$ or $K_3$) by the number of tests carried out for a specific factor in low (medium or high) levels.. For example, $K_1$ for $P$ of $\Delta\beta$ is the sum of $P$ in Table 4 when $\Delta\beta$ is low (i.e., cases 1–3), and $\bar{K}_1$ is obtained by dividing $K_1$ by 3. And the range

$$R = \max\{\bar{K}_1, \bar{K}_2, \bar{K}_3\} - \min\{\bar{K}_1, \bar{K}_2, \bar{K}_3\} \tag{12}$$

$R$ represents the impact of $\Delta\beta$, $D_r$, and $D_d$ on performance ($P$, $\eta$, or $P\eta$). $R$ in Table 5 is plotted as a bar graph in Fig. 4(d). The color of the bar represents different performances, and the height of the bar represents $R$ of each factor; the higher the height is, the greater the impact of the factor. A comparison of $R$ of the different factors in Fig. 4(d) shows that $D_r$ has the largest, $D_d$ the next, and $\Delta\beta$ the smallest impact on performance $P$. This indicates that $D_r$ is the main factor affecting $P$. For $\eta$ ($P\eta$), the values of $R$ for each factor are in the order of $D_d > D_r > \Delta\beta$, which means that the effects of $\Delta\beta$ $D_r$ and $D_d$ on $\eta$ ($P\eta$) gradually increase. In addition, to determine the optimum levels of $\Delta\beta$, $D_r$, and $D_d$, a line graph of $\overline{K}_1$, $\overline{K}_2$, and $\overline{K}_3$ for different performances ($P$, $\eta$, and $P\eta$) of $\Delta\beta$, $D_r$, and $D_d$ is plotted in Fig. 4(a)–(c).

Table 5. Range analysis of $\Delta\beta$, $D_r$, and $D_d$ on $P$, $\eta$, and $P\eta$.

|  | $P$ | | | $\eta$ | | | $P\eta$ | | |
|---|---|---|---|---|---|---|---|---|---|
|  | $\Delta\beta$ | $D_r$ | $D_d$ | $\Delta\beta$ | $D_r$ | $D_d$ | $\Delta\beta$ | $D_r$ | $D_d$ |
| $K_1$ | 0.092 | 0.066 | 0.173 | 1.762 | 2.086 | 2.400 | 0.039 | 0.035 | 0.113 |
| $K_2$ | 0.171 | 0.105 | 0.129 | 2.062 | 1.872 | 1.926 | 0.098 | 0.057 | 0.059 |
| $K_3$ | 0.108 | 0.199 | 0.069 | 1.873 | 1.739 | 1.371 | 0.053 | 0.099 | 0.019 |
| $\overline{K}_1$ | 0.031 | 0.022 | 0.058 | 0.587 | 0.695 | 0.800 | 0.013 | 0.012 | 0.038 |
| $\overline{K}_2$ | 0.057 | 0.035 | 0.043 | 0.687 | 0.624 | 0.642 | 0.033 | 0.019 | 0.020 |
| $\overline{K}_3$ | 0.036 | 0.066 | 0.023 | 0.624 | 0.580 | 0.457 | 0.018 | 0.033 | 0.006 |
| $R$ | 0.026 | 0.044 | 0.035 | 0.100 | 0.116 | 0.343 | 0.020 | 0.022 | 0.031 |

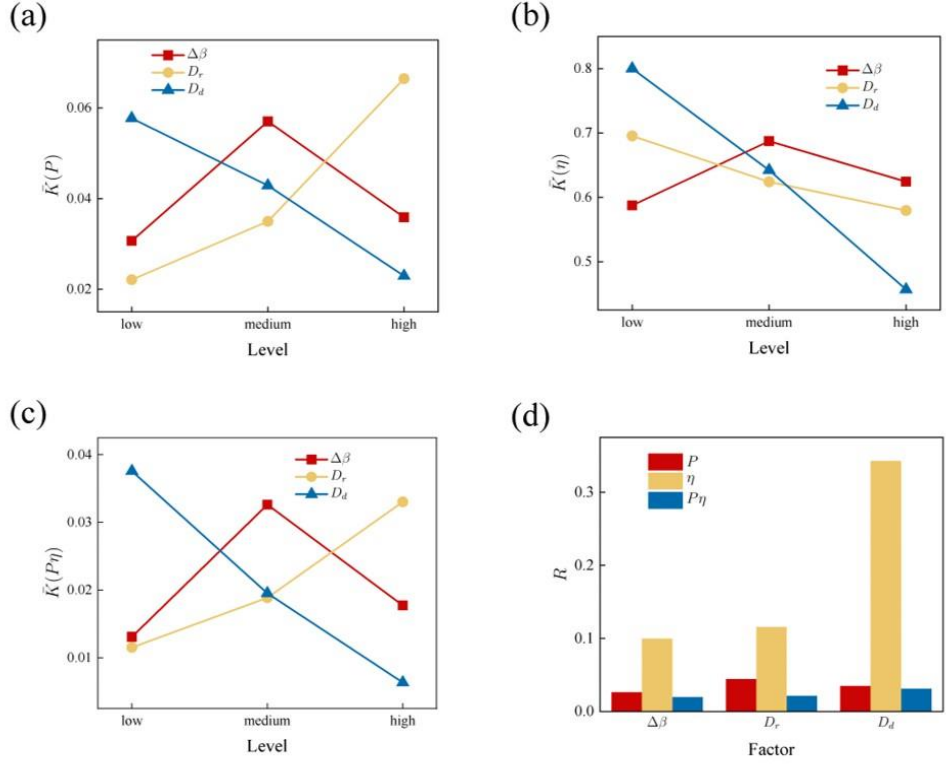

Fig. 4. Range analysis. (a) Line graph of $\bar{K}(P)$ for different levels of $\Delta\beta$, $D_r$, and $D_d$. (b) Line graph of $\bar{K}(\eta)$ for different levels of $\Delta\beta$, $D_r$, and $D_d$. (c) Line graph of $\bar{K}(P\eta)$ for different levels of $\Delta\beta$, $D_r$, and $D_d$. (d) Histogram of $R$ for $\Delta\beta$, $D_r$, and $D_d$ of $P$, $\eta$, and $P\eta$.

From Fig. 4, it can be observed that the $\bar{K}$ values of $\Delta\beta$ in Fig. 4(a), (b), and (c) all reach a maximum at the medium level, which means that the performance is optimal at a medium temperature difference. Meanwhile, the maximum $\bar{K}$ value of $D_r$ is reached at high, low, and high resonant dissipations on $P$, $\eta$, and $P\eta$, respectively., This suggested that high resonant benefits $P$ ($P\eta$), and low resonant benefits $\eta$. The maximum $\bar{K}$ value of $D_d$ in Fig. 4(a), (b), and (c) occurs at a low level, indicating that a lower detuning benefits the performance of the three-level QHE. These results imply that for the best $P$ and $P\eta$, the combinations of $\Delta\beta$, $D_r$, and $D_d$ should be medium, high, and low, respectively. For an efficient $\eta$, the best combination was medium $\Delta\beta$, low $D_r$ and low $D_d$.

### 3.2.2 Analysis of variance

The range analysis indicates the impact of each factor and selects the optimal corresponding levels; however, it cannot indicate whether a factor has a significant effect on the results. Therefore, an additional analysis of variance (ANOVA) was performed. In addition, the impacts of $\Delta\beta$, $D_r$, and $D_d$ on the performance obtained through ANOVA and range analysis should be the same, which verifies the conclusions of the range analysis. The results of the ANOVA on $P$, $\eta$, and $P\eta$ are presented in Table 6.

Table 6. Analysis of variance (ANOVA) for $P$, $\eta$, and $P\eta$ of $\Delta\beta$, $D_r$, and $D_d$.

| Performance | Source of difference | $S$ | $df$ | $\hat{S}$ | $F$ | $p$ |
|---|---|---|---|---|---|---|
| $P$ | $\Delta\beta$ | $1.17\times10^{-3}$ | 2 | $5.85\times10^{-4}$ | 6.814 | 0.128 |
| | $D_r$ | $3.13\times10^{-3}$ | 2 | $1.56\times10^{-3}$ | 18.208 | 0.052 |
| | $D_d$ | $1.83\times10^{-3}$ | 2 | $9.14\times10^{-4}$ | 10.645 | 0.086 |
| $\eta$ | $\Delta\beta$ | $1.53\times10^{-2}$ | 2 | $7.64\times10^{-3}$ | 8.511 | 0.105 |
| | $D_r$ | $2.05\times10^{-2}$ | 2 | $1.02\times10^{-2}$ | 11.399 | 0.081 |
| | $D_d$ | $1.77\times10^{-1}$ | 2 | $8.85\times10^{-2}$ | 98.592 | 0.010* |
| $P\eta$ | $\Delta\beta$ | $6.26\times10^{-4}$ | 2 | $3.13\times10^{-4}$ | 2.945 | 0.253 |
| | $D_r$ | $7.18\times10^{-4}$ | 2 | $3.59\times10^{-4}$ | 3.377 | 0.228 |
| | $D_d$ | $1.48\times10^{-3}$ | 2 | $7.38\times10^{-4}$ | 6.941 | 0.126 |

\* $p < 0.05$ \*\* $p < 0.01$(\* means generally significant, \*\* means more significant)

The first column in Table 6 lists three performance evaluation indexes: $P$, $\eta$, and $P\eta$. The second column lists the different factors ($\Delta\beta$, $D_r$, and $D_d$). $S$, $df$, $\hat{S}$, and $F$ are the sums of the squared deviation, space of freedom, mean square deviation, and ratio of impact [56], respectively. Taking these four parameters into a specific distribution formula yields $p$ [56], which is used to determine whether this factor has a significant impact on the results, where $0.01 < p < 0.05$ (marked by *) indicates that this factor generally has a significant impact, and $p < 0.01$ (marked by **) means that it is more significant [56]. Fig. 5 shows $p$ values of $\Delta\beta$, $D_r$, and $D_d$ on $P$, $\eta$, and $P\eta$ in Table 6. The dashed lines in the graph indicate whether there is a significant effect (generally significant between the two dashed lines, and more significant below the lower dashed line). Fig. 5 shows that only $p$ value of $D_d$ on the $\eta$ is between the two dashed lines, which means that only $D_d$ (i.e., detuning dissipation) has a generally significant impact on the $\eta$.

The value of $p$ also indicates the impact of each factor on the performance (the smaller the $p$, the greater is the impact). From Fig. 5, it can be concluded that the impacts of $\Delta\beta$, $D_r$, and $D_d$ on $P$, $\eta$ or $P\eta$, are consistent with the conclusions of the range analysis in the previous section.

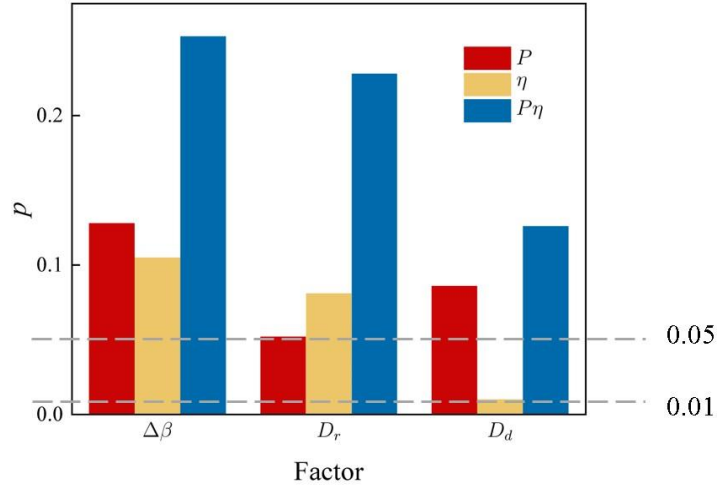

Fig. 5. Histogram of ANOVA of $\Delta\beta$, $D_r$, and $D_d$ on $P$, $\eta$, and $P\eta$. The bar color represents the different performance indicators, whereas the bar height represents the value of $p$ and the horizontal coordinates of the bar represent different factors.

### 3..2.3 Selection of the optimal combination of parameters

The results obtained from the range analysis and ANOVA were consistent, indicating the credibility of the orthogonal test results and conclusions. Therefore, based on these results, the optimal level of each factor was selected in the order of $D_d$, $D_r$, and $\Delta\beta$, as listed in Table 7.

Table 7. The best combination selected based on orthogonal test results.

| Performance | $D_d$ | $D_r$ | $\Delta\beta$ |
|---|---|---|---|
| $P$ | low | high | medium |
| $\eta$ | low | low | medium |
| $P\eta$ | low | high | medium |

As can be observed in Table 7, the optimal level for $\Delta\beta$ is medium and the optimal level for $D_r$ to achieve a higher efficiency is low. These findings are unexcepted. According to the settings in Table 1, medium $\Delta\beta$ represents a medium hot/cold reservoir temperature difference, but the performance should be maximized at a higher $\Delta\beta$. Similarly, $\eta$ should also reach a maximal value under a higher $D_r$. The mechanism behind these phenomena will be further discussed in the next section.

### 3.3 Further analysis of the orthogonal test results

The rate of total heat flow from the hot reservoir in the three-level QHE under the stationary limit can be rewritten as follows:

$$\Phi_h = \frac{1}{\eta^{CP}} P + \rho_0 P_0 \tag{13}$$

where $\eta^{CP}$ is coupling efficiency, and $\rho_0 P_0$ represents the heat leakage from the hot reservoir to the cold reservoir. A higher $\eta^{CP}$ leads to higher performance by reducing the heat losses within the coupling, while a higher $\rho_0 P_0$ increases the useless non-power extraction heat and decreases the performance. For the convenience of analysis, $1/\eta^{CP}$ and $\rho_0 P_0 / P$ were selected as the indicators for analysis, and the smaller they are, the better the performance.

**3.3.1 Higher efficiency in a lower resonant**

First, a study on $D_d$ and $D_r$ under a medium temperature difference (i.e., $\Delta\beta$ is set as the medium level) was conducted to illustrate that the efficiency is maximized at a low $D_r$. The results are shown in Fig. 6. Comparing the middle and bottom rows in Fig. 6, it can be observed that the value of $\eta$ is maximized at low $D_r$, whereas for the top row, the low $D_d$, each column has almost the same $\eta$. Moreover, the efficiency in the top row reaches its maximum value at the boundary, whereas the middle and bottom rows exhibit minimal values on the edge. It can be concluded that the efficiency is sensitive to the dissipation coefficient, $\omega_{20}/\omega_{10}$, and $\lambda/\omega_{10}$. When there is detuning ($D_d$ is not low), a lower resonance (lower $D_r$) benefits the efficiency rather than a higher one.

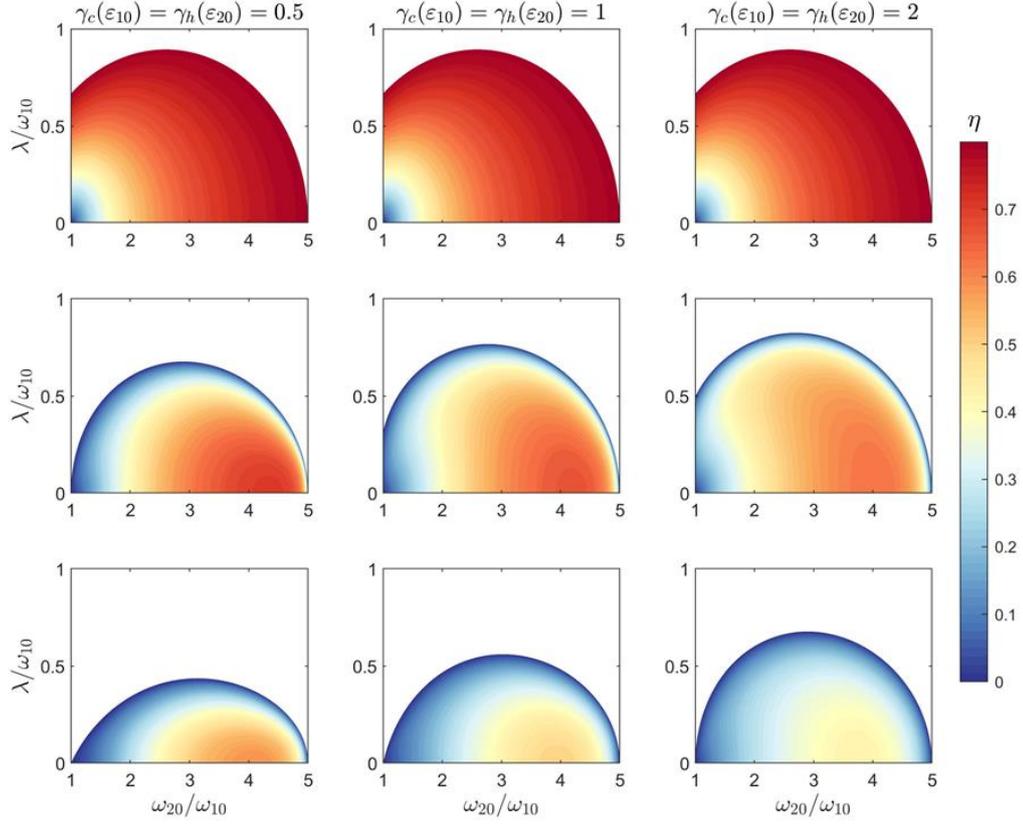

Fig. 6. Efficiency of the QHE in the low, medium, and high dissipation cases under the medium temperature differences. The temperatures of the cold and hot reservoirs are set as $\beta_c\omega_{10}$=2.5, $\beta_h\omega_{10}=0.5$, respectively. The left, middle, and right columns correspond to low, medium, and high levels of $D_r$, whereas the rows from top to bottom represent low, medium, and high $D_d$, respectively.

This phenomenon can be explained by analyzing the rate of total heat flow given in Eq. (13). $1/\eta^{\mathrm{CP}}$ and $\rho_0 P_0/P$ for different dissipations in the medium $\Delta\beta$ are depicted in Fig. 7(a) and (b), respectively. The dashed lines marked with "Low," "Medium," and "High" correspond to low, medium, and high $D_r$ in the orthogonal test, respectively. It can be found from Fig. 7(a) and (b) that both $1/\eta^{\mathrm{CP}}$ and $\rho_0 P_0/P$ on the left are smaller than those on the right when $D_d$ is greater than zero. The larger $D_d$ is, the larger the difference between the left and right sides. Thus, a lower $D_r$ benefits the efficiency by improving the coupling efficiency and reducing the heat leakage from the hot reservoir to the cold reservoir; this benefit becomes more significant when $D_d$ is increased.

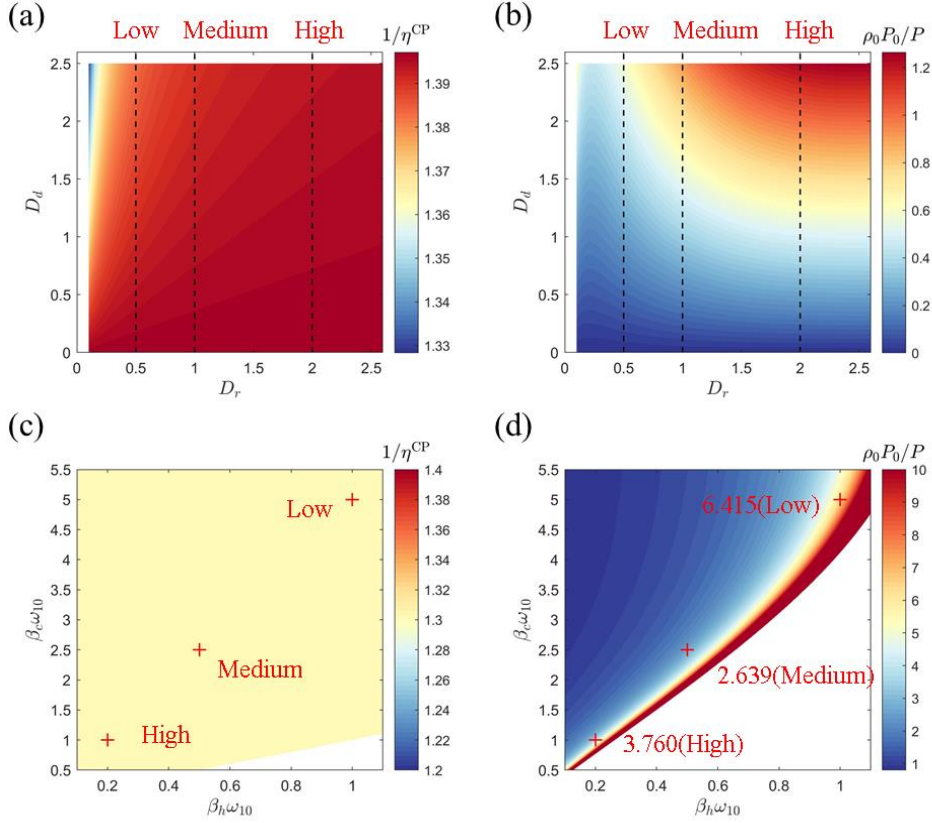

Fig. 7. Components of the rate of total heat flow $\Phi_h$ for different dissipation coefficients ($D_r$ and $D_d$) and different temperature differences ($\Delta\beta$). (a) $1/\eta^{CP}$ and (b) $\rho_0 P_0/P$ for different $D_r$ and $D_d$. The setting of $D_r$ is $\gamma_c(\varepsilon_{10})/\omega_{10} = \gamma_h(\varepsilon_{20})/\omega_{10}$ and takes a value from 0.1 to 2.6. The setting of $D_d$ is $\gamma_c(\varepsilon_{20})/\omega_{10} = \gamma_h(\varepsilon_{10})/\omega_{10}$ and takes a value from 0 to 2.5. The temperatures of the cold and hot reservoirs are $\beta_c\omega_{10}=2.5$, $\beta_h\omega_{10}=0.5$, respectively, and $\omega_{20}/\omega_{10}=3.5$, $\lambda/\omega_{10}=0.1$. (c) $1/\eta^{CP}$ and (d) $\rho_0 P_0/P$ under different $\Delta\beta$ values, where $\beta_h\omega_{10}$ takes values from 0.1 to 1.1 and $\beta_c\omega_{10}$ takes values from 0.5 to 5.5, aimed to avoid the high and low temperature limits. The specific value for $D_r$ is $\gamma_c(\varepsilon_{10})/\omega_{10}=2$, $\gamma_h(\varepsilon_{20})/\omega_{10}=2$ and for $D_d$ is $\gamma_c(\varepsilon_{20})/\omega_{10}=2$, $\gamma_h(\varepsilon_{10})/\omega_{10}=2$; $\omega_{20}/\omega_{10}=2.6$, $\lambda/\omega_{10}=0.5$.

**3.3.2 Higher performance under a medium temperature difference**

A further study on $\Delta\beta$ under the uniform coupling case (i.e., $D_r$ and $D_d$ were both set at a high level) was conducted to explain the results mentioned at the end of Section 3.2. The results are presented in Fig. 8. It can be found that the values of $P$, $\eta$, and $P\eta$ all maximized at medium temperature difference (middle column in Fig. 8). $P$, $\eta$, and $P\eta$ have different sensitivities to the energy level difference $\omega_{20}/\omega_{10}$ and the external field intensity $\lambda/\omega_{10}$ when the temperature difference increases (from the left column to the right column). For $P$, the maximum value was obtained when $\omega_{20}/\omega_{10}$ was small and $\lambda/\omega_{10}$ was relatively high for all the temperature difference cases (low, medium, and high). This is because a higher $\omega_{20}/\omega_{10}$ will improve the theoretical optimal value, but requires a higher field intensity $\lambda/\omega_{10}$ to achieve this value in practice. The best value of $\eta$ was obtained at a larger $\omega_{20}/\omega_{10}$ and zero $\lambda/\omega_{10}$ for all temperature difference cases because a higher $\omega_{20}/\omega_{10}$ will improve the theoretical efficiency regardless of the power, but a higher $\lambda/\omega_{10}$ causes a loss of energy and hence a lower efficiency. For $P\eta$, the maximal value was obtained when $\omega_{20}/\omega_{10}$ and $\lambda/\omega_{10}$ were both medium for all the temperature difference cases because it was the product of $P$ and $\eta$.

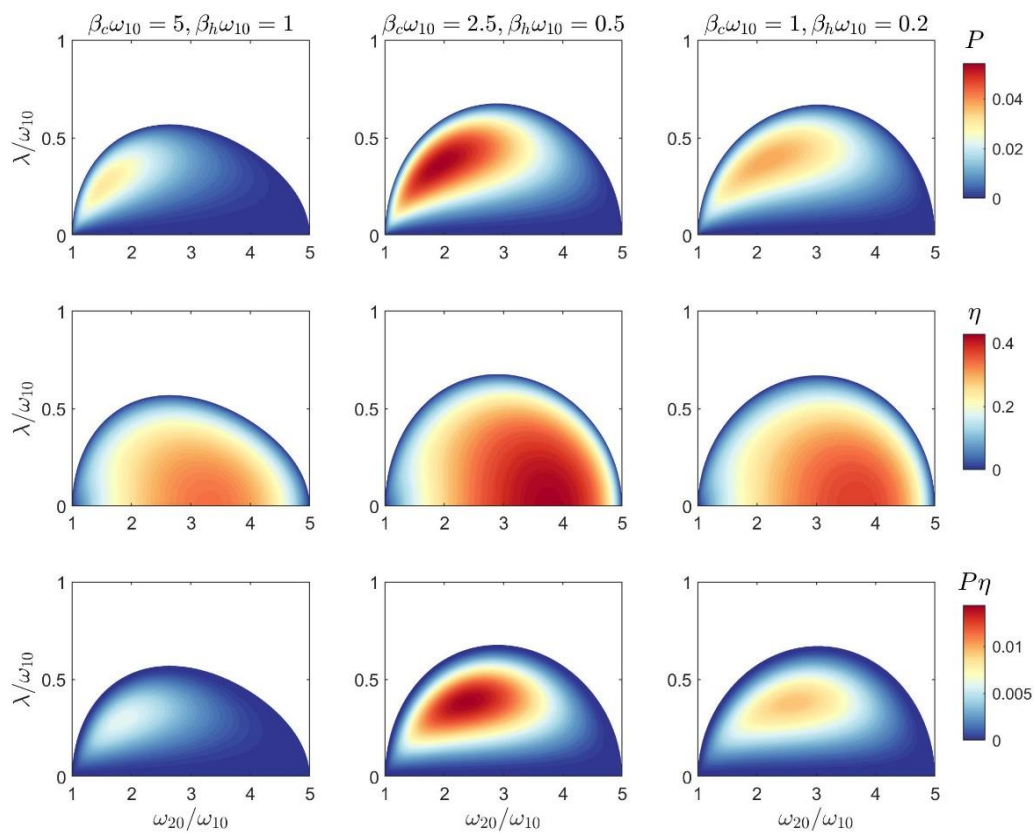

Fig. 8. Performance of the QHE under the low, medium, and high temperature

difference cases for uniform coupling. The dissipation coefficient is set as $\gamma_c(\varepsilon_{10})/\omega_{10}=2$, $\gamma_h(\varepsilon_{20})/\omega_{10}=2$, $\gamma_c(\varepsilon_{20})/\omega_{10}=2$, $\gamma_h(\varepsilon_{10})/\omega_{10}=2$, and the temperature of the cold and hot reservoirs are set as $\beta_c\omega_{10}=5$, $\beta_h\omega_{10}=1$, $\beta_c\omega_{10}=2.5$, $\beta_h\omega_{10}=0.5$, and $\beta_c\omega_{10}=1$, $\beta_h\omega_{10}=0.2$ from the left to right columns, which correspond to low, medium, and high temperature differences. The rows from top to bottom represent a specific performance evaluation index, i.e., efficiency $P$, efficiency $\eta$, and efficacy $P\eta$.

The improvement in the performance for medium temperature differences over high temperature differences can also be explained by analyzing the components of the rate of total heat flow in Eq. (13). The heat flow components for different temperature differences are shown in Fig. 7(c) and (d). The marked red points with text "High," "Medium," and "Low" correspond to high, medium, and low temperature differences, respectively. The reciprocals of the coupling efficiency for different temperature differences are the same (1.309), because

$$\eta^{CP} = \frac{1}{1-q_2-\frac{\varepsilon_{10}}{\varepsilon_{20}}q_1}\left(1-\frac{\varepsilon_{10}}{\varepsilon_{20}}\right) \tag{14}$$

where

$$q_1 = \frac{\gamma_h(\varepsilon_{10})\frac{1-\cos\theta}{2}}{\gamma_c(\varepsilon_{10})\frac{1+\cos\theta}{2}+\gamma_h(\varepsilon_{10})\frac{1-\cos\theta}{2}}$$

$$q_2 = \frac{\gamma_c(\varepsilon_{20})\frac{1-\cos\theta}{2}}{\gamma_h(\varepsilon_{20})\frac{1+\cos\theta}{2}+\gamma_c(\varepsilon_{20})\frac{1-\cos\theta}{2}} \tag{15}$$

and $\varepsilon,\theta$ depend on intensity $\lambda$ and the energy-level level difference $\omega_{20}$. Thus, $1/\eta^{CP}$ does not depend on the temperature, and substituting the values of the parameters into Eq. (14) and Eq. (15) yields 1.309. This implies that the coupling efficiency is constant under different $\Delta\beta$ values as long as the coupling parameters are determined.

For the heat leakage, as shown in Fig. 7(d), the values of $\rho_0 P_0/P$ under low, medium, and high $\Delta\beta$ conditions are 3.760, 2.639, and 6.415, respectively. This implies that the medium $\Delta\beta$ promotes performance by reducing the useless heat leakage.

### 3.4 Discussion on Carnot efficiency and quantum friction

The efficiency of the Carnot cycle is

$$\eta^C = 1 - \frac{T_c}{T_h} = 1 - \frac{\beta_h}{\beta_c} \quad (16)$$

There are three sets of inversed temperatures, $\beta_c\omega_{10}=5$, $\beta_h\omega_{10}=1$ $\beta_c\omega_{10}=2.5$, $\beta_h\omega_{10}=0.5$, and $\beta_c\omega_{10}=1$, $\beta_h\omega_{10}=0.2$. Using Eq. (16), it can be obtained that the Carnot efficiency for all the studied cases is 0.8. The maximum efficiency of the orthogonal test in Table 4 is less than or equal to 0.8, which means that the results are limited by the Carnot efficiency.

The rate of heat flow from the reservoirs to the quantum system $S$ can be calculated using Eq. (1),

$$\dot{Q}_\alpha(t) = \mathrm{tr}\left[\frac{\partial \hat{\rho}_S(t)}{\partial t}\hat{H}_S(t)\right] = \mathrm{tr}\left[\hat{D}[\hat{\rho}_S(t)]\hat{H}_S(t)\right] \quad (17)$$

and $\dot{Q}_\alpha(t)$ can be divided into two parts [57, 58],

$$\dot{Q}_\alpha(t) = \dot{Q}_\alpha^d(t) + \dot{Q}_\alpha^{nd}(t) \quad (18)$$

the diagonal part generated by the change of the level populations

$$\dot{Q}_\alpha^d(t) = \sum_n p_n(t)\hat{D}_\alpha[\hat{\rho}_S(t)]p_n(t)\hat{H}_S(t) \quad (19)$$

and the non-diagonal part related to the energy level coherence

$$\dot{Q}_\alpha^{nd}(t) = \dot{Q}_\alpha(t) - \dot{Q}_\alpha^d(t) \quad (20)$$

where $p_n(t)$ is the population of level $n$. The difference from a classical system is the non-diagonal part induced by the quantum coherence effect. The rational use of the non-diagonal heat can enhance the performance of a QHE. The generation of non-diagonal heat may even consume the internal energy for outputting work resulting in a decreased performance, which is called "quantum friction" [9, 10, 12, 59].

To determine the specific role of "quantum friction," the efficiency is rewritten as diagonal and non-diagonal components

$$\eta = \eta^d \frac{Q_h^d}{Q_h} + \eta^{nd} \frac{Q_h^{nd}}{Q_h} \quad (21)$$

where

$$\eta^{d} = \frac{Q_{h}^{d} + Q_{c}^{d}}{Q_{h}^{d}}$$

$$\eta^{nd} = \frac{Q_{h}^{nd} + Q_{c}^{nd}}{Q_{h}^{d}} \quad (22)$$

In the case of a stationary limit, the population is independent of time $t$, which means that there is an $\dot{Q}^{d}(t) = \dot{Q}_{c}^{d}(t) + \dot{Q}_{h}^{d}(t) = 0$. Therefore, Eq. (21) can be simplified to

$$\eta = \eta^{nd} \frac{Q_{h}^{nd}}{Q_{h}} \quad (23)$$

Meanwhile, the heat engine requires that the output power $P$ be greater than 0, and the rate of heat flow over one cycle $\tau_{0}$ can be rewritten as follows:

$$\Phi_{h}^{nd} = \frac{Q_{h}^{nd}}{\tau_{0}} = \frac{1}{\eta^{nd}} P$$

$$\Phi_{c}^{nd} = \frac{Q_{c}^{nd}}{\tau_{0}} = \left(1 - \frac{1}{\eta^{nd}}\right) P \quad (24)$$

Depending on the value of $1/\eta^{nd}$, the heat flow in a three-level heat engine can be divided into three modes, as presented in Table 8 and Fig. 9. Heat flow mode (1) means that the quantum system absorbs heat from the cold reservoir and releases heat to the heat reservoir; mode (2) means that the quantum system absorbs heat from the heat reservoir and the cold reservoir, whereas mode (3) means that the quantum system absorbs heat from a hot reservoir and releases heat to the cold reservoir. Importantly, with Eq. (24) and the condition that $P$ is greater than or equal to zero, it is not difficult to determine whether both $\Phi_{h}^{nd}$ and $\Phi_{c}^{nd}$ are positive in mode (2). This means that mode (2) is an ideal working mode because all the non-diagonal heat flows are used to produce work and hence, no "quantum friction" exists.

Table 8. Different heat flow modes.

| $1/\eta^{nd}$ | $\Phi_{h}^{nd}$ | $\Phi_{c}^{nd}$ | mode |
| --- | --- | --- | --- |
| $(-\infty, 0)$ | − | + | (1) |
| $(0, 1)$ | + | + | (2) |
| $(1, +\infty)$ | + | − | (3) |

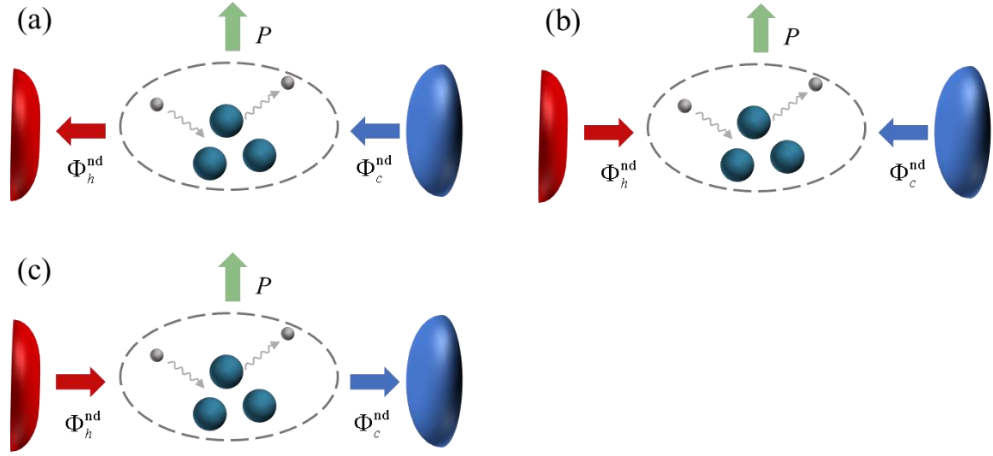

Fig. 9. Different heat flow modes in Table 8. Diagrams (a), (b), and (c) correspond to modes (1), (2), and (3) in Table 8, respectively.

Fig. 10 shows the $1/\eta^{nd}$ of all nine orthogonal test cases for different intensities of the external field $\lambda/\omega_{10}$ and energy level ratio $\omega_{20}/\omega_{10}$. There are five colors in the plots: dark blue, light blue, light yellow, orange, and red. Dark blue indicates that the value of $1/\eta^{nd}$ is less than zero, which implies that the heat flow mode in the heat engine is (1). Light blue and light yellow indicate that the value of $1/\eta^{nd}$ is between zero and one, indicating that the heat flow mode in the heat engine is (2). Orange and red indicate that the value of $1/\eta^{nd}$ is greater than 1, which implies that the heat flow mode in the heat engine is (3). It can be observed from the figure that the values $1/\eta^{nd}$ in cases 1, 6, and 8 are all between 0 and 1 (0.5 to be precise), which means that there is no "quantum friction." This can explain why cases 1, 6, and 8 exhibited the highest efficiency in the orthogonal test results (Table 4). According to the previously designed orthogonal test (Table 2), $D_d$ in cases 1, 6, and 8 is the smallest (zero), which means that "quantum friction" can be avoided and the efficiency can be improved by reducing the detuning dissipation coefficient ($D_d$).

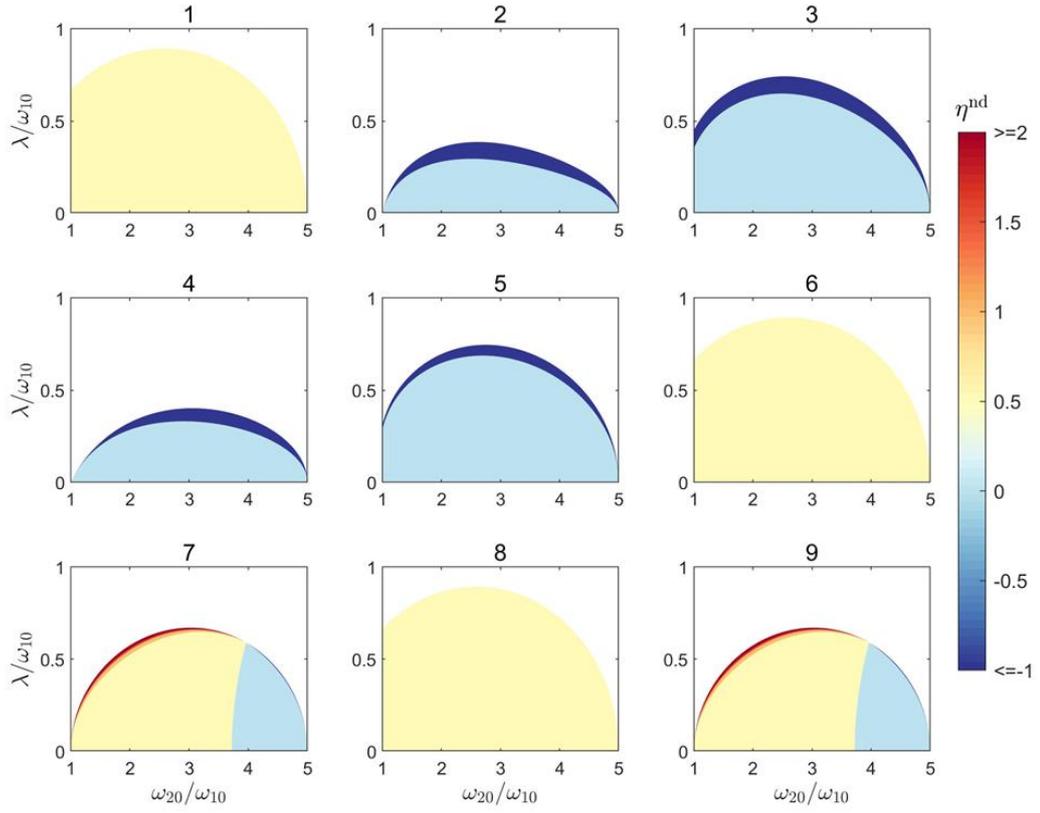

Fig. 10. $1/\eta^{nd}$ of each case in the orthogonal test. The number of each panel indicates the case included in the orthogonal tests. The intensity of the external field $\lambda/\omega_{10}$ ranged from 0 to 1, and the energy-level ratio $\omega_{20}/\omega_{10}$ ranged from 1 to 5 in each case.

## 4 Conclusions

In this study, an orthogonal test was designed to investigate the steady-state performance of a continuous three-level QHE, and the parameters that affected the performance were ordered and optimized by range and variance analyses. The results of the range and variance analyses are explained further. The main conclusions are as follows:

1) The ranking of the parameters affecting the performance of this QHE is $D_d > D_r > \Delta\beta$, that is, detuning dissipation > resonant dissipation > temperature difference.

2) The best-level combination for $D_d$, $D_r$ and $\Delta\beta$ is low, high (low on efficiency), and medium, that is, low detuning, high (low) resonance, and medium temperature difference between the hot and cold reservoirs.

3) The results of the range and variance analyses can be understood from the

perspective of total heat flow losses. The efficiency maximized under a low $D_r$ is due to the reduced nonpower extraction and heat leakage. The medium $\Delta\beta$ decreases the heat leakage, leading to optimal performance.

4) The maximum efficiency was limited by the Carnot efficiency, and the "quantum friction" induced by detuning leads to lower maximal efficiency.

## Acknowledgement

This work was supported by the (Grand No. tsqn202103142), Natural Science Foundation of Shandong Province (No. ZR2021QE033) and China Postdoctoral Science Foundation (No. 2021M702013).

## Data Availability

The data that support the findings of this study are available from the corresponding author upon reasonable request.

## References

[1] H.E.D. Scovil and E.O. Schulz-DuBois, Three-level masers as heat engines, *Phys. Rev. Lett.* **2**, 262 (1959).

[2] J.E. Geusic, E.O. Schulz-DuBios and H.E.D. Scovil, Quantum equivalent of the carnot cycle, *Phys. Rev.* **156**, 343 (1967).

[3] S. Bhattacharjee and A. Dutta, Quantum thermal machines and batteries, *Eur. Phys. J. B* **94**, 1-42 (2021).

[4] H.-T. Quan, Y.-x. Liu, C.-P. Sun and F. Nori, Quantum thermodynamic cycles and quantum heat engines, *Phys. Rev. E* **76**, 031105 (2007).

[5] H.T. Quan, Quantum thermodynamic cycles and quantum heat engines. II, *Phys. Rev. E* **79**, 041129 (2009).

[6] J. Goold, M. Huber, A. Riera, L. Del Rio and P. Skrzypczyk, The role of quantum information in thermodynamics—a topical review, *J. Phys. A: Math. Theor.* **49**, 143001 (2016).

[7] K. Brandner and U. Seifert, Periodic thermodynamics of open quantum systems, *Phys. Rev. E* **93**, 062134 (2016).

[8] S. Vinjanampathy and J. Anders, Quantum thermodynamics, *Contemp. Phys.* **57**, 545-579 (2016).

[9] R. Kosloff and T. Feldmann, Discrete four-stroke quantum heat engine exploring the origin of friction, *Phys. Rev. E* **65**, 055102 (2002).

[10] F. Plastina, A. Alecce, T.J.G. Apollaro, G. Falcone, G. Francica, F. Galve, N.L. Gullo and R. Zambrini, Irreversible work and inner friction in quantum thermodynamic processes, *Phys. Rev. Lett.* **113**, 260601 (2014).

[11] K. Brandner, M. Bauer and U. Seifert, Universal coherence-induced power losses of quantum heat engines in linear response, *Phys. Rev. Lett.* **119**, 170602 (2017).

[12] S. Deng, A. Chenu, P. Diao, F. Li, S. Yu, I. Coulamy, A. Del Campo and H. Wu, Superadiabatic quantum friction suppression in finite-time thermodynamics, *Sci. Adv.* **4**, eaar5909 (2018).

[13] M.O. Scully, K.R. Chapin, K.E. Dorfman, M.B. Kim and A. Svidzinsky, Quantum heat engine power can be increased by noise-induced coherence, *Proc. Natl. Acad. Sci.* **108**, 15097-15100 (2011).

[14] M.T. Mitchison, M.P. Woods, J. Prior and M. Huber, Coherence-assisted single-shot cooling by quantum absorption refrigerators, *New J. Phys.* **17**, 115013 (2015).

[15] C.B. Dağ, W. Niedenzu, Ö.E. Müstecaplıoğlu and G. Kurizki, Multiatom quantum coherences in micromasers as fuel for thermal and nonthermal machines, *Entropy* **18**, 244 (2016).

[16] J. Roßnagel, O. Abah, F. Schmidt-Kaler, K. Singer and E. Lutz, Nanoscale heat engine beyond the Carnot limit, *Phys. Rev. Lett.* **112**, 030602 (2014).

[17] G. Manzano, F. Galve, R. Zambrini and J.M.R. Parrondo, Entropy production and thermodynamic power of the squeezed thermal reservoir, *Phys. Rev. E* **93**, 052120 (2016).

[18] W. Niedenzu, V. Mukherjee, A. Ghosh, A.G. Kofman and G. Kurizki, Quantum engine efficiency bound beyond the second law of thermodynamics, *Nat. Comm.* **9**, 1-13 (2018).

[19] K.V. Hovhannisyan, M. Perarnau-Llobet, M. Huber and A. Acín, Entanglement generation is not necessary for optimal work extraction, *Phys. Rev. Lett.* **111**,

240401 (2013).

[20] M. Perarnau-Llobet, K.V. Hovhannisyan, M. Huber, P. Skrzypczyk, N. Brunner and A. Acín, Extractable work from correlations, *Phys. Rev. X* **5**, 041011 (2015).

[21] B. Karimi and J.P. Pekola, Otto refrigerator based on a superconducting qubit: Classical and quantum performance, *Phys. Rev. B* **94**, 184503 (2016).

[22] J.P. Pekola, B. Karimi, G. Thomas and D.V. Averin, Supremacy of incoherent sudden cycles, *Phys. Rev. B* **100**, 085405 (2019).

[23] R. Uzdin, A. Levy and R. Kosloff, Equivalence of quantum heat machines, and quantum-thermodynamic signatures, *Phys. Rev. X* **5**, 031044 (2015).

[24] G. Manzano, F. Plastina and R. Zambrini, Optimal work extraction and thermodynamics of quantum measurements and correlations, *Phys. Rev. Lett.* **121**, 120602 (2018).

[25] S. Seah, S. Nimmrichter and V. Scarani, Maxwell's lesser demon: A quantum engine driven by pointer measurements, *Phys. Rev. Lett.* **124**, 100603 (2020).

[26] Y. Zou, Y. Jiang, Y. Mei, X. Guo and S. Du, Quantum heat engine using electromagnetically induced transparency, *Phys. Rev. Lett.* **119**, 050602 (2017).

[27] Q. Bouton, J. Nettersheim, S. Burgardt, D. Adam, E. Lutz and A. Widera, A quantum heat engine driven by atomic collisions, *Nat. Comm.* **12**, 1-7 (2021).

[28] J. Roßnagel, S.T. Dawkins, K.N. Tolazzi, O. Abah, E. Lutz, F. Schmidt-Kaler and K. Singer, A single-atom heat engine, *Science* **352**, 325-329 (2016).

[29] G. Maslennikov, S. Ding, R. Hablützel, J. Gan, A. Roulet, S. Nimmrichter, J. Dai, V. Scarani and D. Matsukevich, Quantum absorption refrigerator with trapped ions, *Nat. Comm.* **10**, 1-8 (2019).

[30] N. Van Horne, D. Yum, T. Dutta, P. Hänggi, J. Gong, D. Poletti and M. Mukherjee, Single-atom energy-conversion device with a quantum load, *npj Quantum Inform.* **6**, 1-9 (2020).

[31] J. Klatzow, et al., Experimental demonstration of quantum effects in the operation of microscopic heat engines, *Phys. Rev. Lett.* **122**, 110601 (2019).

[32] K. Ono, S.N. Shevchenko, T. Mori, S. Moriyama and F. Nori, Analog of a quantum heat engine using a single-spin qubit, *Phys. Rev. Lett.* **125**, 166802 (2020).

[33] W. Ji, Z. Chai, M. Wang, Y. Guo, X. Rong, F. Shi, C. Ren, Y. Wang and J. Du, Spin Quantum Heat Engine Quantified by Quantum Steering, *Phys. Rev. Lett.* **128**, 090602 (2022).

[34] G.L. Zanin, M. Antesberger, M.J. Jacquet, P.H.S. Ribeiro, L.A. Rozema and P. Walther, Enhanced Photonic Maxwell's Demon with Correlated Baths, *arXiv preprint arXiv:2107.09686* (2021).

[35] J.V. Koski, V.F. Maisi, T. Sagawa and J.P. Pekola, Experimental Observation of the Role of Mutual Information in the Nonequilibrium Dynamics of a Maxwell Demon, *Phys. Rev. Lett.* **113**, 030601 (2014).

[36] V. Koski Jonne, F. Maisi Ville, P. Pekola Jukka and V. Averin Dmitri, Experimental realization of a Szilard engine with a single electron, *Proc. Natl. Acad. Sci.* **111**, 13786-13789 (2014).

[37] J.V. Koski, A. Kutvonen, I.M. Khaymovich, T. Ala-Nissila and J.P. Pekola, On-Chip Maxwell's Demon as an Information-Powered Refrigerator, *Phys. Rev. Lett.* **115**, 260602 (2015).

[38] G. Manzano, D. Subero, O. Maillet, R. Fazio, J.P. Pekola and É. Roldán, Thermodynamics of Gambling Demons, *Phys. Rev. Lett.* **126**, 080603 (2021).

[39] P.A. Camati, J.P.S. Peterson, T.B. Batalhão, K. Micadei, A.M. Souza, R.S. Sarthour, I.S. Oliveira and R.M. Serra, Experimental Rectification of Entropy Production by Maxwell's Demon in a Quantum System, *Phys. Rev. Lett.* **117**, 240502 (2016).

[40] J.P.S. Peterson, T.B. Batalhão, M. Herrera, A.M. Souza, R.S. Sarthour, I.S. Oliveira and R.M. Serra, Experimental Characterization of a Spin Quantum Heat Engine, *Phys. Rev. Lett.* **123**, 240601 (2019).

[41] R.J. de Assis, T.M. de Mendonça, C.J. Villas-Boas, A.M. de Souza, R.S. Sarthour, I.S. Oliveira and N.G. de Almeida, Efficiency of a Quantum Otto Heat Engine Operating under a Reservoir at Effective Negative Temperatures, *Phys. Rev. Lett.*

**122**, 240602 (2019).

[42] N. Cottet, et al., Observing a quantum Maxwell demon at work, *Proc. Natl. Acad. Sci.* **114**, 7561-7564 (2017).

[43] L. Bresque, P.A. Camati, S. Rogers, K. Murch, A.N. Jordan and A. Auffèves, Two-Qubit Engine Fueled by Entanglement and Local Measurements, *Phys. Rev. Lett.* **126**, 120605 (2021).

[44] P. Bayona-Pena and K. Takahashi, Thermodynamics of a continuous quantum heat engine: Interplay between population and coherence, *Phys. Rev. A* **104**, 042203 (2021).

[45] E. Geva and R. Kosloff, Three-level quantum amplifier as a heat engine: A study in finite-time thermodynamics, *Phys. Rev. E* **49**, 3903-3918 (1994).

[46] N. Gunantara, A review of multi-objective optimization: Methods and its applications, *Cogent Eng.* **5**, 1502242 (2018).

[47] S. Katoch, S.S. Chauhan and V. Kumar, A review on genetic algorithm: past, present, and future, *Multimed. Tools Appl.* **80**, 8091-8126 (2021).

[48] B. Li, Z. Cui, Q. Cao and W. Shao, Increasing Efficiency of a Finned Heat Sink Using Orthogonal Analysis, *Energies* **14**, 782 (2021).

[49] R. Uzdin, The second law and beyond in microscopic quantum setups in *Thermodynamics in the Quantum Regime: Fundamental Aspects and New Directions*, pp. 681-712 (Springer, 2018).

[50] A. Levy and R. Kosloff, The local approach to quantum transport may violate the second law of thermodynamics, *Europhys. Lett.* **107**, 20004 (2014).

[51] F. Brandao, M. Horodecki, N. Ng, J. Oppenheim and S. Wehner, The second laws of quantum thermodynamics, *Proc. Natl. Acad. Sci.* **112**, 3275-3279 (2015).

[52] P. Strasberg and A. Winter, First and second law of quantum thermodynamics: a consistent derivation based on a microscopic definition of entropy, *PRX Quantum* **2**, 030202 (2021).

[53] R. Uzdin and S. Rahav, Passivity deformation approach for the thermodynamics

of isolated quantum setups, *PRX Quantum* **2**, 010336 (2021).

[54] R. Uzdin and S. Rahav, Global passivity in microscopic thermodynamics, *Phys. Rev. X* **8**, 021064 (2018).

[55] R. Uzdin, Additional energy-information relations in thermodynamics of small systems, *Phys. Rev. E* **96**, 032128 (2017).

[56] D.C. Montgomery, *Design and analysis of experiments* (John wiley & sons, 2017).

[57] J.P. Santos, L.C. Céleri, G.T. Landi and M. Paternostro, The role of quantum coherence in non-equilibrium entropy production, *npj Quantum Inform.* **5**, 23 (2019).

[58] T. Baumgratz, M. Cramer and M.B. Plenio, Quantifying coherence, *Phys. Rev. Lett.* **113**, 140401 (2014).

[59] A.I. Volokitin and B.N.J. Persson, Quantum friction, *Phys. Rev. Lett.* **106**, 094502 (2011).